\documentclass[twocolumn]{aastex6}
\usepackage{amsmath,ulem,textcomp}
\usepackage{color,soul}
\usepackage[flushleft]{threeparttable}
\newcommand{\nuclei}[2]{$^{#1}${#2}}
\newcommand{\pgreaction}[4]{\nuclei{#1}{#2}($p,\gamma$)\nuclei{#3}{#4}}

\newcommand{\hedreaction}[4]{\nuclei{#1}{#2}($^3$He$,d$)\nuclei{#3}{#4}}
\newcommand{\abundance}[1]{[{#1}/Fe]}
\begin{document}

\title{Sensitivity to thermonuclear reaction rates in modeling the abundance anomalies of NGC 2419}
\author{J.R. Dermigny \altaffilmark{1,2} and C. Iliadis \altaffilmark{1,2}}
\altaffiltext{1}{Department of Physics \& Astronomy, University of North Carolina, Chapel Hill, NC 27599-3255, USA}
\altaffiltext{2}{Triangle Universities Nuclear Laboratory, Durham, NC 27708-0308, USA}
\email{dermigny@unc.edu}
\begin{abstract}
Abundance anomalies in globular clusters provide strong evidence for multiple stellar populations within each cluster. 
These populations are usually interpreted as distinct generations, with the currently observed second-generation stars having formed in part
from the ejecta of massive, first-generation ``polluter" stars, giving rise to the anomalous abundance patterns. The precise nature of the polluters and their enrichment mechanism are still unclear.
Even so, the chemical abundances measured in second-generation stars within the globular cluster NGC 2419 provide insight into this puzzling process.
Previous work used Monte Carlo nuclear reaction network calculations to constrain the temperature-density conditions that could reproduce the observed abundances, thereby placing robust limits on the origins of the polluter material. The effect of individual reaction rates on these conditions has not been studied, however.
Thus, we perform an exhaustive sensitivity study on the nuclear physics input to determine which reactions have the greatest impact on these predictions. 
We find that the \nuclei{30}{Si}(p,$\gamma$)\nuclei{31}{P}, \nuclei{37}{Ar}(p,$\gamma$)\nuclei{38}{K}, \nuclei{38}{Ar}(p,$\gamma$)\nuclei{39}{K},
and \nuclei{39}{K}(p,$\gamma$)\nuclei{40}{Ca} reactions are all critical in determining the temperature-density conditions, and ultimately, the origin of the polluter material. We conclude with recommendations for future experiments.
\end{abstract}


\section{Introduction} \label{sec:intro}
Many aspects of globular cluster formation and evolution are unknown. These stellar aggregates were once thought to house only a single stellar population, however, recent photometric analysis suggests that they instead harbor multiple, chemically distinct populations. This is indicated by the presence of discrete tracks in a color-magnitude diagram \citep{Villanova_2007,Piotto_2007} and significant star-to-star variations in the abundances of light elements (e.g., C, N, O, Na, Mg, Al). 
These signatures cannot be attributed to a single stellar generation formed from inhomogeneous interstellar matter.
Therefore, theories for globular cluster formation now assume multiple \textit{generations} of stars, i.e., first and second generation \citep{D_Ercole_2008,Carretta_2010}.
The second generation is thought to form from the highly processed ejecta of the first-generation ``polluter" stars, giving rise to the abundance variations observed today. 
The details of this self-enrichment scenario, including the nature of the polluters, are still unclear.
\par
NGC 2419 is a massive, metal-poor globular cluster in the Milky Way outer halo which is now emblematic of the self-enrichment phenomenon. 
Photometric analysis of this cluster by \citet{Cohen_2011}, \citet{Cohen_2012}, and \citet{Mucciarelli_2012} revealed a population of red giants with an unprecedented depletion in magnesium ([Mg/Fe] $\approx -1$)\footnote{According to common convention, abundances are given as $[A/B]\equiv \log_{10}(N_A/N_B)_\star - \log_{10}(N_A/N_B)_\odot$, where $N_i$ are number abundances of elements $A$ and $B$ observed in a star $(\star)$ or the sun $(\odot)$; while the quantity $[A/B]$ is unitless, differences between two values are expressed in units of dex (``decimal exponent").}. 
This depletion is coupled to a strong enhancement in potassium ([K/Fe] $\approx 2$), giving rise to a Mg-K anticorrelation within NGC 2419. 
The origin of this anticorrelation is not yet clear.   
Of particular interest are the evolutionary time-scales involved, the identity of the polluter stars, and the enrichment mechanism.
Once understood, these characteristics may shed new light on globular cluster and galaxy formation as well as provide a means to distinguish between competing models of self-enrichment.
\par
With this potentially substantial impact on our understanding of globular clusters, enrichment scenarios for NGC 2419 have been studied extensively. First, \citet{Ventura_2012} explored the possibility that the severe Mg depletion was caused by high temperature burning within first-generation asymptotic giant branch (AGB) and super asymptotic giant branch (super AGB) stars. In their model, the second generation formed directly from the ejecta of the first-generation stars. They found that the chemical abundances of the enriched material could be made to agree with the observations provided both model parameters and thermonuclear reaction rates were fine-tuned. Later, \citet{Iliadis_2016} broadened the scope of this investigation. They used Monte Carlo nuclear reaction network calculations to constrain the physical conditions that could give rise to the observed abundances. They found a correlation between stellar temperature and density values that provided a satisfactory match between simulated and observed abundances. In comparing these conditions with stellar models, they found that low-mass stars, AGB stars, massive stars, and super-massive stars are unlikely to account for the abundance anomalies. Super AGB stars and classical novae, involving either carbon-oxygen or oxygen-neon white dwarfs, could be the polluters. 
However, the authors emphasized the need for improved stellar models before drawing firm conclusions.
\par
The thermonuclear reaction rates used to carry out these studies are based on experimental measurements and theoretical calculations. 
Some rates are only known to within a factor of $10$, owing to the uncertainty in their nuclear physics input. 
Therefore, reaction rate uncertainties must be considered for the most robust predictions. 
\citet{Iliadis_2016} already demonstrated that these uncertainties affect the range of plausible polluter sites. 
They found that the allowed region of the temperature-density parameter space widens considerably when the impact of reaction rate uncertainties are considered.
In this work, we determine which reactions contribute to this broadening, with the ultimate goal of constraining the parameter space.
In doing so, we aim to refine the list of polluter candidates.
\par
In Section 2, we describe the nucleosynthesis model used to study this anomalous globular cluster.
We then explore the effect of varying the reaction rate uncertainties and identify reactions critical to reproducing the observed abundances in Section 3.
In Section 4, we review the nuclear physics input to reaction rate calculations. Recommendations for future experiments follow. Finally, a summary is given in Section 5.
\section{Methods} \label{sec:methods}
In this work, we adopt the methodology of \citet{Iliadis_2016} to explore possible self-enrichment scenarios in NGC 2419. For a thorough exposition, we refer the reader to that work. Below, we review the salient features of our model.
\subsection{Polluter Model} \label{sec:methos1}
\par
The adopted model aims to reproduce the observed abundances of NGC 2419 by simulating its formation history. This comprises three major episodes. First, the first stars form following primordial nucleosynthesis. These stars contribute metals to the proto-cluster gas, bringing the overall metallicity to the currently observed value of $[\text{Fe}/\text{H}]= -2.09 \pm .02$ \citep{Mucciarelli_2012}. The first-generation globular cluster stars then form from this material. 
Later, the most massive of these stars will evolve quickly and explode as type II supernovae. 
Others will evolve more slowly, giving rise to the anomalous chemical enrichment. 
Sometime during their evolution, these first-generation stars eject part of their matter into the intracluster gas.
Finally, the currently observed second-generation stars form from this polluted material, thereby inheriting the nucleosynthetic signatures of both pre-enrichment material and the first-generation stars. 
\par
The situation described above is reproduced in our simulations, albeit with simplifications. 
For instance, for the composition of the proto-cluster gas, we adopted the results of a one-zone chemical evolution model for the Milky Way halo. This model
is an updated version of \citet{Goswami_2000}, which reproduces, with some minor adjustments, the reported abundances of field stars of the same average metallicity as NGC 2419 (see the Appendix of \citet{Iliadis_2016} for details).
Further, to keep our results as general as possible, we did not adopt any particular stellar evolution model. Instead, we used a nuclear reaction network to process stellar material in a single zone with constant temperature and density.
\par
The nuclear reaction network follows the evolution of 213 nuclides, ranging from p, n, \nuclei{4}{He}, to \nuclei{55}{Cr}. We adopted the thermonuclear reaction rates linking these nuclides (2373 total) from STARLIB \citep{Sallaska_2013}. This library provides the rate
probability density function at temperatures from $10$ MK to $10$ GK. This unique feature will be discussed further in Section~\ref{sec:stats}.
\par
The simulated stellar evolution occurs as follows. First, the proto-cluster, or ``pristine" gas, is processed at constant temperature, $T$, and density, $\rho$,  using the nuclear reaction network. The evolution is halted after a variable amount of hydrogen, $\Delta X^H\equiv X_{i}^H-X_{f}^H$, has been consumed. This ``processed" matter corresponds to the composition of the first-generation stars before they eject material back into the cluster. For the second-generation stars, we assume that the processed material is diluted with $f$ parts of pristine, intra-cluster gas, as given by:
\begin{equation}
X_{mix} = \frac{X_{proc} + f X_{pris}}{1+f}
\end{equation}
where $X_{proc}$ and $X_{pris}$ denote the mass fractions of the reaction network output (i.e., processed matter) and the initial composition (i.e., pristine matter), respectively. 
The elemental abundances of $X_{mix}$ (i.e., Mg, K, Ca, Ti, Si, Sc, and V) are then compared with those of the observed, extreme population. 
For these stars, we adopt the same abundance conditions as \cite{Iliadis_2016}, which is based on the scatter in the data and abundance uncertainties of individual stars. These are given by 1.3 $<$ \abundance{K} $<$ 2.0, -1.5 $<$ \abundance{Mg} $<$ -0.8, 0.1 $<$ \abundance{Ca} $<$ 0.7, -0.2 $<$ \abundance{Ti} $<$ 0.7, 0.4 $<$ \abundance{Si} $<$ 1.1, 0.4 $<$ \abundance{Sc} $<$ 1.3, and -0.2 $<$ \abundance{V} $<$ 0.6.
If there is agreement for any value of $f$, then the physical conditions used in the simulation represent plausible conditions for the nucleosynthesis of the polluter material. 

\par
Since this approach is so simple, we must exercise caution in interpreting the results. An important limitation of our simulation is that hydrodynamical mechanisms such as convection are absent. In a realistic scenario, the burning region is continuously mixed with fresh (unprocessed) material, replenishing the hydrogen available for burning while diluting the byproducts. Compared with the single-zone process adopted in this work, it would require a much longer burning period to achieve the same abundance levels. For this reason, any constraints we obtain on the amount of hydrogen consumed or equivalently, the duration of burning, are not meaningful. Instead, we focus on the temperature and density, since these parameters determine which nucleosynthesis pathways are operating, and ultimately, the elements synthesized. By repeating this simulation many times, the accumulated temperature and density conditions provide valuable constraints on the origin of this material.

\subsection{Reaction Rate Uncertainties}
\label{sec:stats}
In our simulations, the elemental abundances of the polluter material depend on the interplay between thousands of reactions.
\citet{Iliadis_2016} showed that when reaction rate uncertainties were taken into account, 
the range of physical conditions that could reproduce the observed abundances broadened significantly.
This implies that, for one or more reactions, the rate uncertainty is large enough to have a substantive effect on the outcome of the simulation. 
Identifying these reactions will lead to more robust theories for the origin of the polluter material, as these important reactions
can then be carefully studied in nuclear physics laboratories.
\par
We study the influence of both the statistical and systematic contributions to the reaction rate uncertainties.
The uncertainties adopted by STARLIB comprise the statistical component and any known systematic effects.
These vary widely from reaction to reaction, owing to differences in the mechanism of the reaction (i.e., non-resonant versus resonant reaction) as well as the availability of experimental data.
For the majority of the reactions relevant to this study, the rates have been derived using the experimental Monte Carlo formalism described in \cite{Longland_2010}. Using this method, the uncertainty is determined by considering statistical and known systematic uncertainties of the nuclear physics input, e.g., resonance energies and strengths, partial widths, and resonant S-factors, from physically motivated probability densities. 
Experimental Monte Carlo rates are available for only 64 reactions in the target mass range from A~$=14$ to A~$=40$. 
For reaction rates taken from other sources, such as the NACRE compilation \citep{Angulo_1999}, uncertainties are evaluated on a case-by-case basis. 
Theoretical Hauser-Fechbach rates are adopted from the TALYS computer program \citep{Konig_2004} in the instances where there are no reliable, experimentally determined reaction rates. For these rates, STARLIB adopts a conservative factor uncertainty of $10$.
\par
What is of main interest is not the magnitude of these rate uncertainties but rather the knowledge of the full probability distributions.
Accordingly, for each reaction STARLIB provides two temperature-dependent quantities, $\mu$ and $\sigma$, which describe the rate probability density as a log-normal distribution,
\begin{equation}
f(x) = \frac{1}{\sigma\sqrt{2\pi}}\frac{1}{x}e^{-(\ln x-\mu)^2/(2\sigma^2)},
\end{equation}
where $\mu$ and $\sigma$ determine the location and width of the distribution. 
To sample the reaction rates, we follow the procedure detailed in \citet{Iliadis_2015}.
First, a random number, $p_i$, is generated from a Gaussian distribution with a mean of zero and standard deviation of unity. Sample $i$ of reaction rate $y$ is then given by
\begin{equation}
\label{eq:samp}
y_i = y_{med}(f.u.)^{p_i},
\end{equation}
where $y_{med}$ and $f.u.$ are the median value and the factor uncertainty. These quantities are given by $e^\mu$ and $e^\sigma$, respectively \citep{Longland_2012}. 
\par
We must also consider the possibility that \textit{unknown} systematic errors exist.
This would be the case, for example, if there was an unknown resonance contributing to the reaction.
Additional possibilities include errors in experimental procedure (e.g., target preparation, charge collection) or poor choice of a nuclear model. 
These effects cannot easily be anticipated. 
Nevertheless, we perform tests to explore the possibility that the temperature and density predictions are sensitive to their presence. 
To do so, we first sample the rate according to Eq.~\ref{eq:samp}. 
We then apply a multiplicative variation factor, $\alpha$, such that the new rate is given by $\alpha \times y_i$. 
This has the effect of scaling the probability distribution of this reaction rate by a factor $\alpha$. 
In the next section, we use both of these methods as a means to explore the rate sensitivities in our nuclear reaction network calculations.
\section{Results} \label{sec:methods}
\begin{figure}
\centering{
\plotone{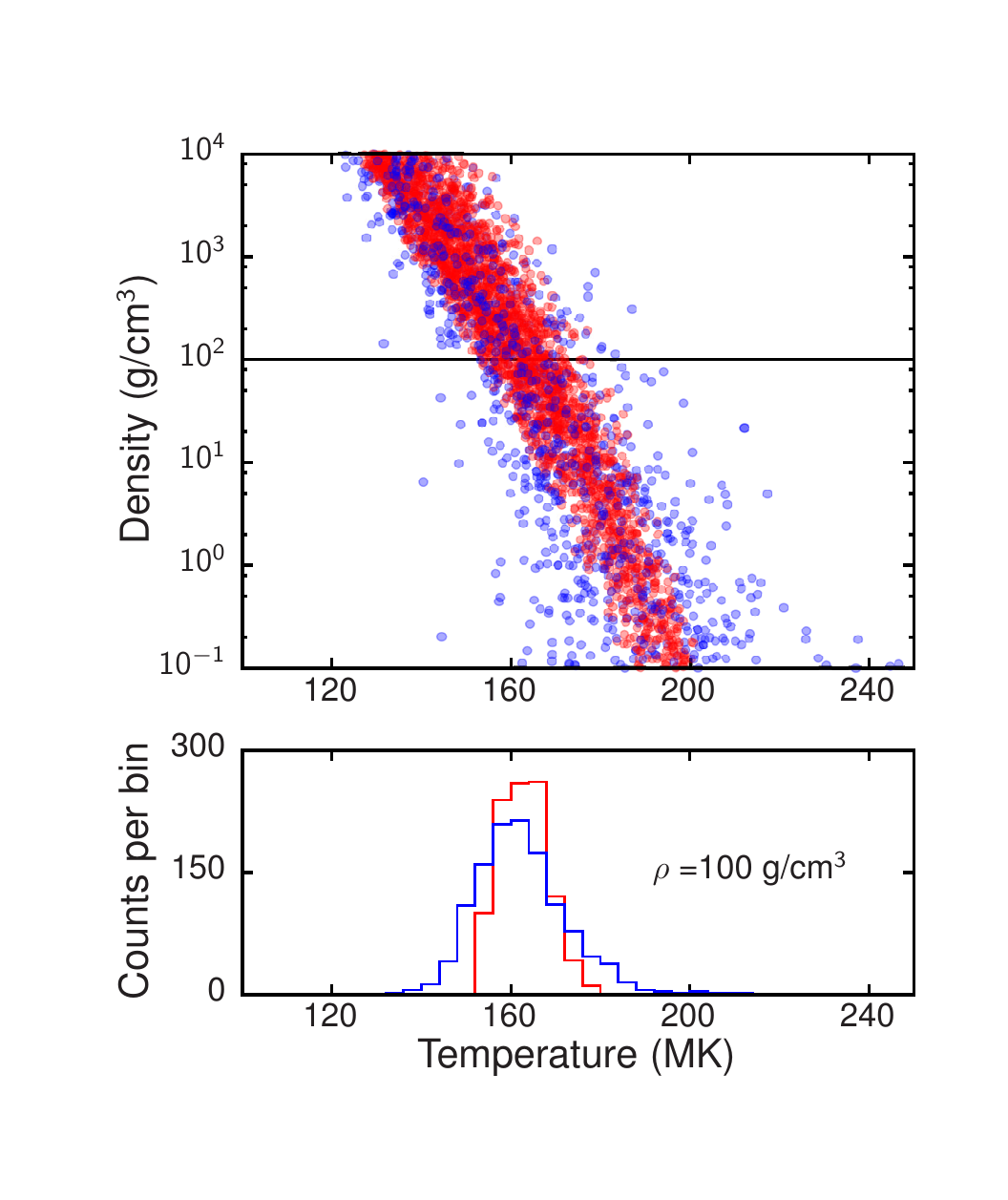}
}
\figcaption{ (Top) Constant temperature-density conditions that reproduce measured abundances in NGC 2419. Red dots result from runs with fixed reaction rates, while blue dots result from runs with varying rates. (Bottom) The projection of acceptable solutions onto the temperature axis. Note that the red histogram has been scaled to $30\%$ of its original height for comparison purposes.}
\label{fig:dots}
\end{figure}

We begin by reproducing the results of \citet{Iliadis_2016}, as they serve as a convenient baseline for comparison.
We adopted their recommended initial composition and assumed fixed thermonuclear rates (i.e., disregarding any uncertainties). For each network simulation, we varied the temperature, $T$, density, $\rho$, and final hydrogen abundance, $X^H_f$, using a Monte Carlo routine. We performed $1 \times 10^5$ of these network simulations, twice the amount used in the earlier work, to ensure that the parameter space was thoroughly sampled.
At the conclusion of each one, the mixtures of the ``pristine" and ``processed" matter were compared against the observed abundances for mixing fractions of $f=1000$ (i.e., almost purely pristine matter), $100$, $30$, $10$, $3$, $1.0$, $0.1$, $0.05$, $0.02$, and $0.0$ (i.e., purely processed matter). For conditions that gave a satisfactory fit to the abundances, the parameters are plotted as red dots in the top panel of Figure~\ref{fig:dots}. 
The width of this distribution indicates, for a given density, the range of temperatures capable of producing the abundances of all the relevant elements in NGC 2419.  
\par
Next, we repeat this procedure while also varying the reaction rates according to their rate probability distribution in STARLIB. Each reaction rate in the network was sampled using Eq.~\ref{eq:samp}. In the case of correlated reactions, i.e., forward and reverse reaction pairs, the rates were sampled together in order to preserve reciprocity.
The results are plotted as blue dots in Figure~\ref{fig:dots}. 
The degree of broadening due to the reaction rate uncertainties is substantial, particularly at low densities. To assess this effect quantitatively, these simulations were repeated with a fixed density of $\rho=100 \text{ g}/\text{cm}^3$. Those that reproduced the measured abundances were then binned according to their temperature. These are shown in the bottom panel of Figure~\ref{fig:dots}. Using this representation, we define the degree of broadening as $\delta$, the difference in temperature between the $16$th and $84$th percentiles of the distribution. At this density, the cumulative effect of varying all reaction rates results in $\delta_{\text{vr}}(100 \text{ g/cm}^3)=19.0$ MK. When expressed as a factor increase over the results obtained with the recommended rates (i.e., their median values), which gives $\delta_{\text{mr}}(100 \text{ g/cm}^3)=11.0$ MK, we find that $\delta_{\text{vr}}/\delta_{\text{mr}}=1.73$. This number indicates the increase in scatter of the temperature values at fixed density. To determine which of the individual reaction rate uncertainties have the greatest impact on the nucleosynthesis, we now consider the broadening due to each reaction. 
\par
\par
\begin{figure}[hbt!]
\centering{
\plotone{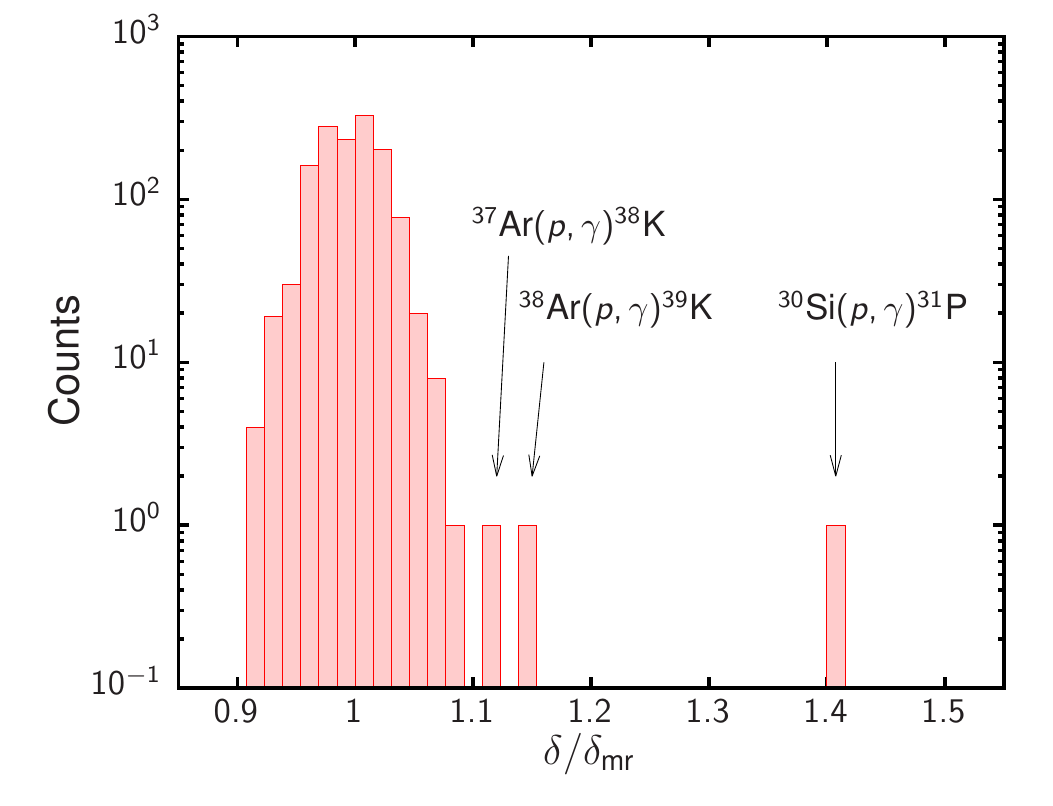}
}
\figcaption{Broadening of the (fixed-density) temperature conditions, $\delta/\delta_{\text{mr}}$, due to the uncertainty of each reaction in our network, as listed in STARLIB. Note that the values for \pgreaction{37}{Ar}{38}{K}, \pgreaction{38}{Ar}{39}{K}, and \pgreaction{30}{Si}{31}{P} lie to the far-right of the centroid.}
\label{fig:histo}
\end{figure}
\par
For each reaction in our network, we performed $2 \times 10^4$ simulations where we sampled \textit{only} the corresponding reaction rate. All other reaction rates remained fixed at their recommended values. The parameters $T$ and $X_f^H$ were varied as before, while $\rho$ remained fixed at $100 \text{ g}/\text{cm}^3$. 
From the accepted samples, the temperature values were then used to measure the broadening for each reaction.

Note that the number of simulations performed was reduced to make this calculation more tractable.
To ensure that these results were not affected by this reduction, this process was repeated for a select few reactions using $1 \times 10^5$ simulations, a fivefold increase. For each of these reactions, the higher statistics results were in agreement to within $3\%$, demonstrating that $2 \times 10^4$ network simulations per reaction is sufficient.
The distribution of all the measured $\delta/\delta_{\text{mr}}$ values, the factor increase over the result obtained using the recommended (median) rates, is shown in Figure~\ref{fig:histo}. 
They appear normally distributed with a mean of $0.99$ and standard deviation of $0.03$. This suggests that the majority of the reaction rates are known to sufficient accuracy with regard to the nature of the polluter material, since their uncertainties have a negligible contribution to the broadening.
There are three outliers, with values of $\delta/\delta_{\text{mr}}=1.12$, $1.15$ and $1.41$, corresponding to the \pgreaction{37}{Ar}{38}{K},  \pgreaction{38}{Ar}{39}{K}, and \pgreaction{30}{Si}{31}{P} reactions, respectively. Their significant influence on the width of the temperature distribution demonstrates that these reactions play a prominent role in determining which stellar conditions, i.e., temperature and density, are capable of reproducing the observed abundances. 
\par
Next, we explore whether the allowed $T-\rho$ conditions are sensitive to unknown systematic effects, i.e., those not included in STARLIB. In addition to the three reactions identified above, we also included in this test those which are important to the synthesis and destruction of both magnesium and potassium. These include the reactions comprising the Ne-Na and Mg-Al nucleosynthesis chains \citep{Rolfs_1988}, as well as the Ar-K chain ~\citep{Iliadis_2016} and the reaction, \pgreaction{39}{K}{40}{Ca}.
For each of these reactions, we performed a series of $1 \times 10^5$ network simulations where their corresponding rates were given an artificial systematic effect.  
The rates of these reactions were varied in two steps. First, they were sampled within their STARLIB uncertainty. 
Each rate sample was then multiplied by the fixed variation factor, $\alpha$, representing the contribution of some unknown effect (see Section~\ref{sec:stats}). 
The factors considered here were $1/10$, $1/5$, $5$, and $10$. 
All other reactions were varied within their STARLIB uncertainties.
\begin{figure*}
\centering{
\plotone{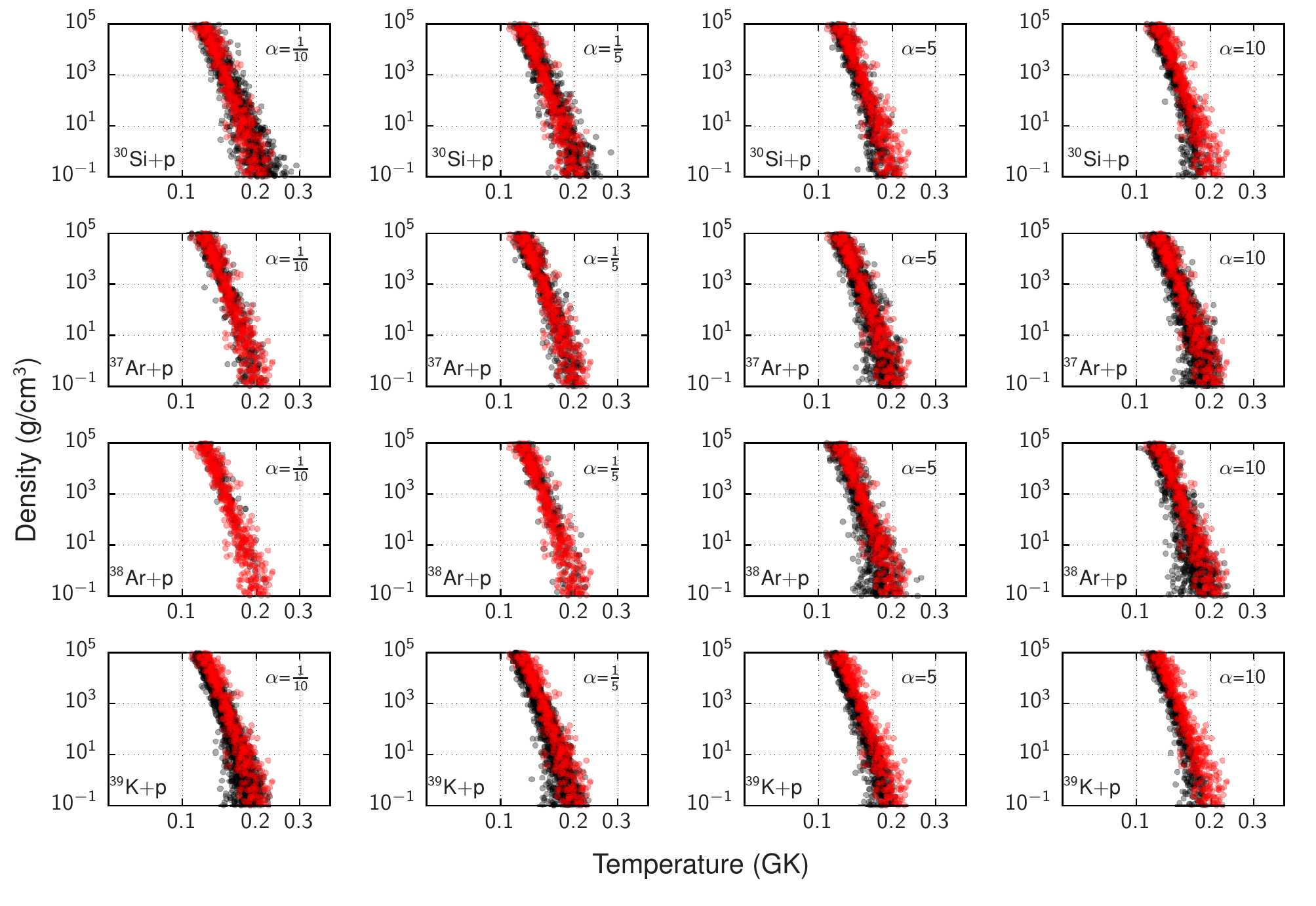}
}
\figcaption{Sensitivity of the temperature-density conditions to the influence of unknown systematic effects in the reaction \pgreaction{30}{Si}{31}{P} (first row), \pgreaction{37}{Ar}{38}{K} (second row), \pgreaction{38}{Ar}{39}{K} (third row), and \pgreaction{39}{K}{40}{Ca} (fourth row). The variation factors ($\alpha=1/10$, $1/5$, $5$, $10$) applied to each reaction rate are shown increasing from left to right. The temperature and density combinations that provide an acceptable match between simulated and observed abundances are shown as black dots. The case with no artificial systematic effect ($\alpha=1$) is shown as red dots, for comparison.}
\label{fig:systematics}
\end{figure*}
\par
Figure~\ref{fig:systematics} shows the sampled parameter space for each simulation, where the rows correspond (from top to bottom) to the \pgreaction{30}{Si}{31}{P}, \pgreaction{37}{Ar}{38}{K}, \pgreaction{38}{Ar}{39}{K}, and \pgreaction{39}{K}{40}{Ca} reactions. For all of the other reactions considered, the effect of the variation was found to be much less impactful. 
The variation factor $\alpha$ is displayed in the top right-hand corner of each plot. The case with no artificial systematic effect, i.e., $\alpha=1$, is shown in the foreground (red dots) for comparison.
The effect on the broadening of the temperature and density conditions (black dots) reveals their sensitivity to the magnitude of each reaction rate. 
For instance, when the \pgreaction{30}{Si}{31}{P} reaction rate is increased by a factor of 5, we find that the $T-\rho$ scatter becomes narrower, with the high-temperature edge receding at any given density. The simulated abundances at these conditions (e.g., $T=170$ MK, $\rho=100$ g/cm$^3$, $X_f^H=0.50$) are found to be depleted in silicon when compared with the $\alpha=1$ case, with a net reduction of $\approx 1.3$ dex. This indicates that the narrowing is caused by the onset of silicon depletion via \pgreaction{30}{Si}{31}{P}. 
\par
Adjustments to the \pgreaction{37}{Ar}{38}{K}, \pgreaction{38}{Ar}{39}{K}, and \pgreaction{39}{K}{40}{Ca} reaction rates reveal a similar effect on the potassium abundance.
Note that as both the \pgreaction{37}{Ar}{38}{K} and \pgreaction{38}{Ar}{39}{K} reaction rates are increased, the $T-\rho$ scatter increases, with newly viable conditions appearing on the low-temperature side. For the \pgreaction{38}{Ar}{39}{K} reaction, the difference between the leftmost and rightmost plot is particularly stark. Under the same test conditions as before, the elemental potassium abundance increases by $\approx 1.0$ dex when this reaction is varied from $\alpha=1/10$ to $10$ times its recommended value. The succeeding reaction, \pgreaction{39}{K}{40}{Ca}, has the opposite effect of depleting potassium by $\approx 0.8$ dex. This is made apparent by comparing the $\alpha = 1/10$ and $10$ cases, where it can be seen that the high-temperature conditions no longer satisfy the abundance constraints when the rate is increased. 
\par
These results demonstrate that systematic effects in these key reaction rates affect the predicted temperature and density conditions that forged the polluter material.
Finally, we explore the possibility that all four of these reactions have unknown contributions and look at their combined effect on these conditions.
We also consider how this affects which potential polluter sites (e.g., classical novae, AGB stars, super AGB stars) are thought to reproduce the observed abundances.
A complete discussion of these candidates is given in \citet{Iliadis_2016}. Their findings are summarized in Section~\ref{sec:intro}.
\par
\begin{figure}
\centering{
\plotone{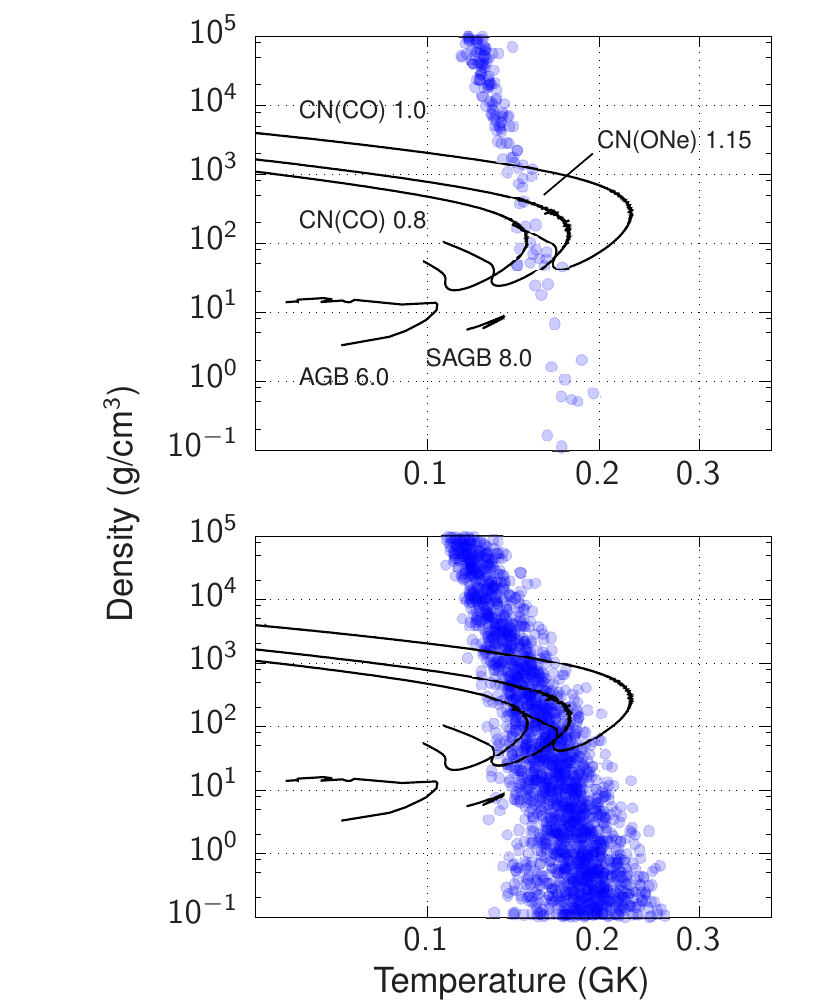}
}
\figcaption{ Temperature and density combinations that provide an acceptable match between simulated and observed abundances are shown as black dots superimposed over temperature-density tracks for carbon-oxygen and oxygen-neon classical novae (``CN") \citep{Iliadis_2016}, AGB \citep{Karakas_2010}, and super AGB \citep{Doherty_2014} models.
(Top) The \pgreaction{30}{Si}{31}{P}, \pgreaction{37}{Ar}{38}{K}, \pgreaction{38}{Ar}{39}{K}, and \pgreaction{39}{K}{40}{Ca} reactions rates have been multiplied by $5$, $1/5$, $1/5$, and $5$, respectively. (Bottom) The variation factors are  $1/5$, $5$, $5$, and $1/5$ respectively.}
\label{fig:extreme}
\end{figure}

We simulated two scenarios. First, we assumed that the \pgreaction{30}{Si}{31}{P}, \pgreaction{37}{Ar}{38}{K}, \pgreaction{38}{Ar}{39}{K}, and \pgreaction{39}{K}{40}{Ca} reaction rates each have variation factors of $\alpha=5$, $1/5$, $1/5$, and $5$, respectively. These factors were chosen to correspond to the case of decreased production of both silicon and potassium.
We then repeated this run, where instead we used factors of $\alpha=1/5$, $5$, $5$, and $1/5$, respectively, representing the opposite situation.
All reaction rates were also varied within their STARLIB uncertainties. 
The results of these two scenarios are shown in Figure~\ref{fig:extreme}, superimposed on the predicted temperature-density evolutions for various stellar models. 
For the variation factors used in the top panel (corresponding to the silicon-poor, potassium-poor scenario) we find that the $T-\rho$ scatter has been drastically reduced. This is especially evident when compared with the bottom panel, where the scatter is much broader, extending over a region $100$ MK wide at lower densities. 
\par
In order to place the most stringent constraints on the origin of the polluter material, minimal scatter in the acceptable temperature and density conditions is preferable. Since this scatter is dependent on the \pgreaction{30}{Si}{31}{P}, \pgreaction{37}{Ar}{38}{K}, \pgreaction{38}{Ar}{39}{K}, and \pgreaction{39}{K}{40}{Ca} reaction rates, our conclusion is that these reactions are critically important to identifying the origin of the polluter material. In the next section, we review the nuclear input data used in the calculation of their thermonuclear rates.
\par
\section{Nuclear Reaction Rates} \label{sec:rates}
The total thermonuclear rate (in units of cm$^3$ mol$^{-1}$ s$^{-1}$) for a 
reaction involving two nuclei ($0$ and $1$) in the entrance channel at a given temperature is given by
\begin{equation}
\label{eq:rate}
\begin{split}
N_A\langle\sigma v \rangle_{01}=\frac{3.7318\times10^{10}}{T_9^{3/2}}\sqrt{\frac{M_0+M_1}{M_0M_1}}
\\ \times \int_{0}^\infty E \sigma(E) e^{-11.605 E/T_9} dE
\end{split}
\end{equation}
where the center-of-mass energy, $E$, is in units of MeV, the temperature, $T_9$, is in GK ($T_9 \equiv T/10^9$ K), the atomic masses, $M_i$, are in u, and the cross section, $\sigma$, is in barn ($1 \text{ b} \equiv 10^{-24}$ cm$^2$), and $N_A$ denotes Avogadro's constant. The cross section is determined by resonant and non-resonant contributions to the reaction. Specialized formulations of Eq.~\ref{eq:rate} have been derived to treat these separately. These can be found, for example, in \citet{Iliadis_Text} and are not repeated here.
\par    
For the \pgreaction{37}{Ar}{38}{K}, \pgreaction{38}{Ar}{39}{K}, \pgreaction{30}{Si}{31}{P}, and \pgreaction{39}{K}{40}{Ca} reactions, the cross section is dominated by narrow, isolated resonances. In this case, the one-level Breit-Wigner formula can be substituted into Eq.~\ref{eq:rate}. This integral can be solved numerically if all partial widths (e.g., $\Gamma_p$, $\Gamma_\alpha$, $\Gamma_\gamma$) are known for a given resonance. Frequently, resonances are so narrow that their cross section cannot be obtained experimentally. Instead, all that is measured is the resonance strength, $\omega\gamma$, which is proportional to the resonance integral.
When using Eq.~\ref{eq:rate}, uncertainties in the experimental nuclear physics data, such as the resonance location and strength, introduce a combined uncertainty to the reaction rate. 
In the following, we explore the nuclear physics input that gives rise to the estimated total uncertainty listed in STARLIB for
these four reactions.
\subsection{$^{37}Ar(p,\gamma)^{38}K$ and $^{38}Ar(p,\gamma)^{39}K$}
\label{sec:argon}
We found that the potassium enhancement was sensitive to the proton capture rates for \nuclei{37}{Ar} and \nuclei{38}{Ar}. 
These reactions, along with several others, comprise a nucleosynthesis chain for potassium production.
This chain begins with proton capture on \nuclei{36}{Ar} and proceeds to the stable isotope \nuclei{39}{K} via the reaction sequence \pgreaction{36}{Ar}{37}{K}$(\beta^+\nu)$\pgreaction{37}{Ar}{38}{K}$(\beta^+\nu)$\pgreaction{38}{Ar}{39}{K}. 
We are particularly interested in these two reaction rates at the temperatures investigated in Section~\ref{sec:methods}, where this reaction sequence is the predominant producer of potassium. These temperatures range from $120$ to $240$ MK, with the corresponding center-of-mass energies of the interacting nuclei ranging from $100$ to $500$ keV. 
\par
First, we will consider the \pgreaction{37}{Ar}{38}{K} reaction.
The target nuclide, \nuclei{37}{Ar}, is a short-lived cosmogenic isotope, with a half-life of $t_{1/2}=35.001(19)$ days \citep{Cameron_2012}. A measurement of this reaction via either proton (a direct measurement) or {\nuclei{3}{He}} bombardment (an indirect measurement) is prohibitively difficult. An attractive alternative is to measure the kinematic inverse, {\nuclei{1}{H}(\nuclei{37}{Ar},$\gamma$)\nuclei{38}{K}}, using radioactive ion beam methods, however, no such experiment has been attempted yet. Since there is scant empirical data on the resonances in this reaction, STARLIB adopts theoretical estimates based on the nuclear reaction model code, TALYS \citep{Konig_2004,Goriely_2008}, for the reaction rate. TALYS calculates nuclear cross sections using experimental information when available. Otherwise, various local and global input parameters are used to represent the nuclear structure properties, optical potentials, level densities, $\gamma$-ray strengths, and fission properties. A comparison between experimental and TALYS cross sections provided in \citet{Sallaska_2013} suggests that the discrepancies for low-mass nuclei (A $< 50$) could be as large as a factor of 10.
While an experimentally determined rate would certainly be preferable, we recommend the use of the TALYS rate until more reaction data is made available.
\par 
Next, we will look at the available nuclear data for the \pgreaction{38}{Ar}{39}{K} reaction.
The strengths and energies of 99 resonances, ranging from the lowest, directly observed resonance at E$_R^{lab}=896.5$ keV to E$_R^{lab}=2400.6$ keV, were measured by \citet{H_nninen_1984}. 
They also calculated the first experimentally derived thermonuclear rate for this reaction. 
This rate was intended for use in modeling explosive oxygen burning, and consequently it was only applicable to temperatures in the range of $T =0.7-4.7$ GK. Later, this rate was re-evaluated for STARLIB by \citet{Sallaska_2013} using the Monte Carlo reaction rate formalism presented in \citet{Longland_2010}. 
This was the first experimental rate calculation for the full range of stellar temperatures.
\par
The STARLIB rate incorporates the effect of several resonances in the $100-500$ keV range.
The corresponding compound levels have been identified using indirect reactions and are listed in Table~\ref{tab:ar_states} along with their available measured properties. 
Experimental data on the level properties for these states is limited. This is evidenced by the ambiguous spin-parity assignments and also the scarcity of partial width measurements.  For instance, the proton partial width, $\Gamma_p$, has only been derived for the state at E$_x=6548$ keV. The $\gamma$-ray partial width has been determined for several other states via lifetime measurements \citep{Nolan_1981,Moreh_1988}. However, without the complementary proton partial width, the precise contribution of these levels to the reaction rate is unknown. 
\par
\begin{table}
\centering
\caption{\pgreaction{38}{Ar}{39}{K} Reaction Data}
\begin{threeparttable}
\makebox[0.85\linewidth]{
    \begin{tabular}{c c c c c} \hline
    E$_{\text{R}}^{cm}$ \tnote{a} & E$_{\text{x}}$ & 2J$^\pi$ & $\Gamma_p$ (eV) & $\Gamma_\gamma$ (eV) \tnote{c} \\ \hline
    120(2) & 6501(2) & $(3,5)^+$ & & $0.0077(28)$ \\
    147(2) & 6528(2) & $(1-7^+)$ & & $0.0050(33)$ \\
    167(2) & 6548(2) & $7^-$     & $7.5 (3.8)\times10^{-13}$\tnote{b} & \\
    272(2) & 6653(2) & $(3,5)^+$ & & $0.019(11)$ \\
    305(2) & 6686(2) &           & & \\
    359(2) & 6740(2) & $(3,5)^+$ & & $0.19(4)$\tnote{d}\\
    437(3) & 6818(3) & $(3,5)^+$ & & $0.14(4)$\tnote{d}\\
    447(2) & 6828(2) &           & & \\
    \hline 
    \end{tabular}}
    \tablecomments{
    Nuclear data on low-energy resonances in the \pgreaction{38}{Ar}{39}{K} reaction. 
    The spin-parity assignments and partial widths are listed for each resonance, when available.
}
    \begin{tablenotes}
        \item[a] The information shown for each of these states is used \newline in the rate evaluation in \citet{Sallaska_2013}.
        \item[b] Based on spectroscopic factors \citep{Kn_pfle_1974}
        \item[c] From \cite{Nolan_1981}, unless otherwise stated.
        \item[d] From \cite{Moreh_1988}.
    \end{tablenotes}

\end{threeparttable}
\label{tab:ar_states}
\end{table}
\par
\begin{figure}[hbt]
\centering{
\includegraphics[width=.4\textwidth]{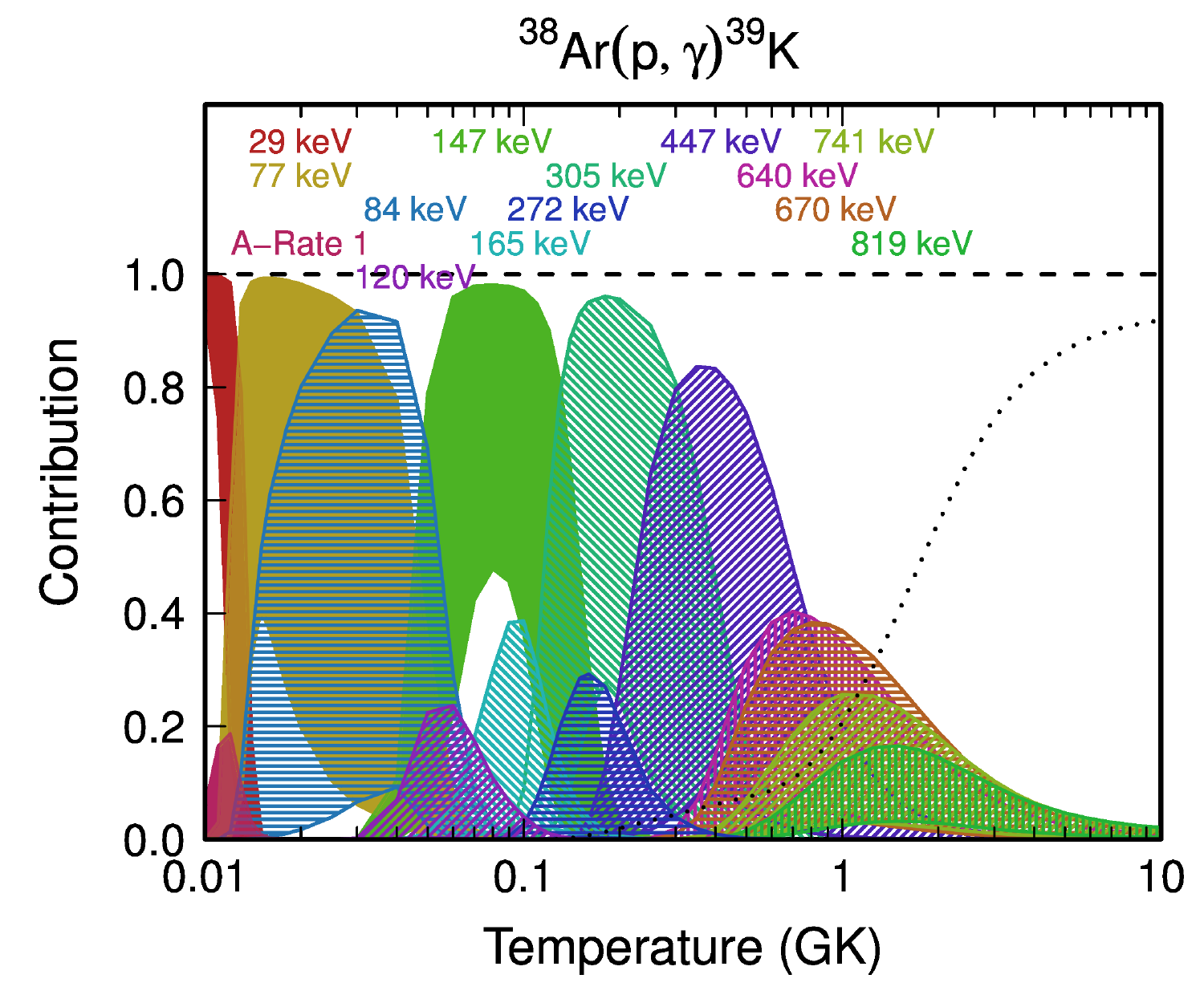} \\
\includegraphics[width=.4\textwidth]{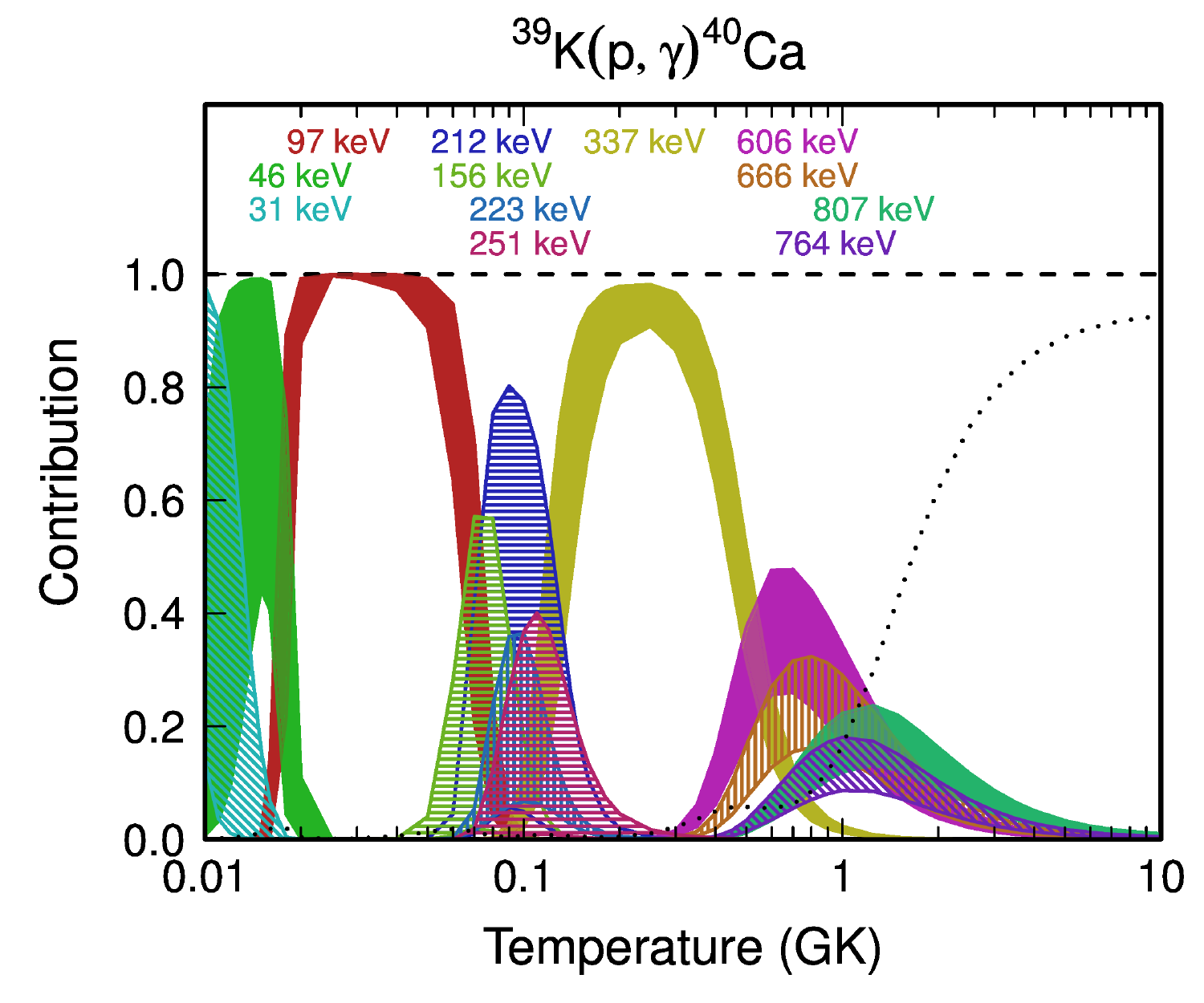} \\
\includegraphics[width=.4\textwidth]{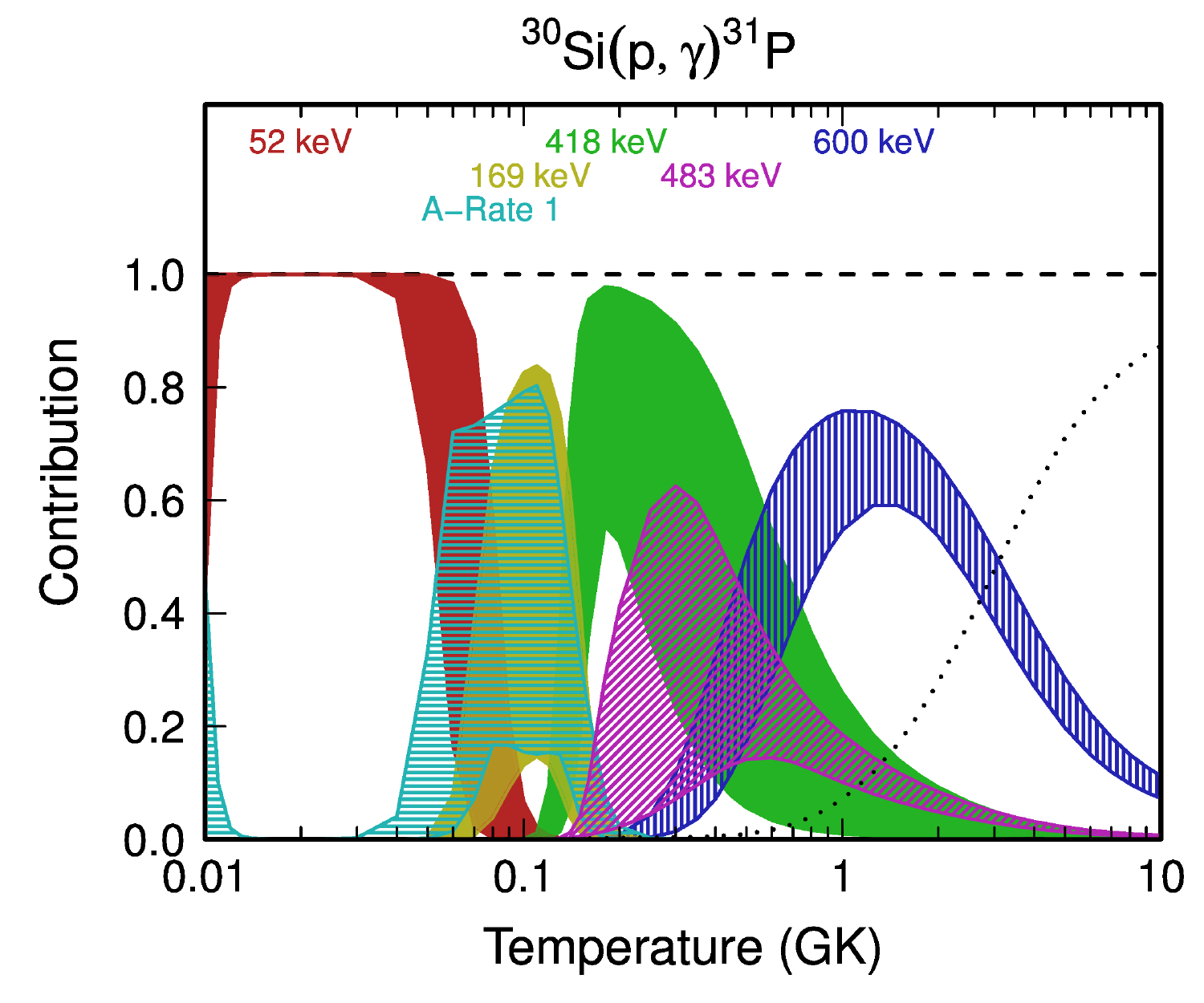}
}
\figcaption{Individual contributions from resonances and non-resonant amplitudes to the total reaction rate at given temperatures for the reactions \pgreaction{38}{Ar}{39}{K} (top), \pgreaction{39}{K}{40}{Ca} (middle), and \pgreaction{30}{Si}{31}{P} (bottom). Contributions were calculated using the STARLIB reaction rates. The vertical width of each band indicates the Monte Carlo uncertainty of the contribution at a fixed temperature and corresponds to a coverage probability of 68\%. Numbers at the top denote center-of-mass energies of given resonances. 
`A-Rate 1' refers to the non-resonant (direct capture) rate contribution. The dotted line at higher temperatures indicates the aggregate contribution of higher lying resonances. }
\label{fig:38ar}
\end{figure}

The details of the reaction rate calculation based on these data are available in \citet{Sallaska_2013}, and we refer the reader to that work. 
We briefly point out that the experimental Monte Carlo reaction rate formalism takes uncertainties in the nuclear physics input into account by sampling the parameters from appropriate probability density functions. 
This unique feature is demonstrated for the \pgreaction{38}{Ar}{39}{K} reaction in the top panel of Figure~\ref{fig:38ar}.
The fractional contribution of each resonance to the thermonuclear reaction rate is presented, where the width of the bands corresponds to 
the Monte Carlo uncertainty of each contribution. 
Between $100$ and $300$ MK, we find that the rate is dominated by contributions from the E$_R^{cm}=147,\ 305, \text{ and } 447$ keV resonances. 
All three exhibit a large degree of uncertainty owing to the lack of experimental data for their proton partial widths (Table~\ref{tab:ar_states}).
Therefore, new measurements would significantly reduce the uncertainties of the \pgreaction{38}{Ar}{39}{K} reaction rate.
\par
\par
\subsection{$^{39}K(p,\gamma)^{38}Ca$}
The potassium synthesized by the Ar-K nucleosynthesis chain is susceptible to destruction through the \pgreaction{39}{K}{40}{Ca} reaction. In Section~\ref{sec:methods}, we found that varying the thermonuclear rate for this reaction produced a marked change in the range of the plausible temperature and density conditions of the polluter material. This effect warrants a closer look at the nuclear physics input to the reaction rate.
\par
The rate adopted in this work is based on the recent evaluation presented in \citet{Longland_2016}, which uses the Monte Carlo reaction rate formalism of \cite{Longland_2010}. For a comprehensive overview of this reaction and the available nuclear data, we refer the reader to this work and the references therein. We note here that the lowest-lying, directly observed resonance is at E$_R^{lab}=622.2$ keV \citep{Kikstra_1990}. Information on the compound levels between $100$ and $500$ keV is therefore limited to indirect measurements. Of the 15 resonances observed in this range, the proton partial width, $\Gamma_p$, has been estimated only for the E$_R^{cm}=97$, $223$, and $337$ keV resonances \citep{Seth_1967,Cage_1971,Fuchs_1969}. Their estimated relative contributions to the reaction rate are illustrated in the middle panel of Figure~\ref{fig:38ar} alongside other important resonances. The key feature of this plot is that the $337$-keV resonance (solid green band) is the main contributer to the $^{39}$K+p reaction at temperatures in the range of $140-400$ MK. Since most of the allowed temperature-density conditions fall within this range (see Figure~\ref{fig:dots}), the final elemental potassium abundance in our simulations is largely determined by the width of this single resonance. Therefore, in addition to the resonances suggested for \pgreaction{38}{Ar}{39}{K} reaction, we also recommend a direct measurement of the E$_{R}^{cm}=337$ keV resonance in \pgreaction{39}{K}{40}{Ca} to further our understanding of the observed potassium enrichment in NGC 2419.
\par
\subsection{$^{30}Si(p,\gamma)^{31}P$}
Silicon synthesis and destruction varies precipitously in the $100-200$ MK temperature range. At the lower-end ($< 140$ MK), silicon is primarily made up of the isotope \nuclei{28}{Si}. In this regime, destruction via proton capture is sufficiently slow, allowing the elemental abundance of silicon to increase steadily by way of the \pgreaction{27}{Al}{28}{Si} reaction. At higher temperatures, \nuclei{28}{Si} is more efficiently converted to \nuclei{30}{Si} via the sequence \pgreaction{28}{Si}{29}{P}$(\beta^+\nu)$\pgreaction{29}{Si}{30}{P}$(\beta^+\nu)^{30}$Si.
At temperatures above about $160$ MK, the accumulated \nuclei{30}{Si} is rapidly consumed via the \pgreaction{30}{Si}{31}{P} reaction, reducing the silicon abundance to sub-solar values.
By varying this rate in Section~\ref{sec:methods}, we found that the onset of this depletion affects the simulated temperature and density conditions of the polluter material.
For this reason, a more precise estimation of the \pgreaction{30}{Si}{31}{P} reaction rate is necessary. 
\par
The most recent experimental rate was published in \citet{Longland_2010} using the Monte Carlo reaction rate formalism.
For stellar temperatures between $T=120-240$ MK, the rate is determined by five resonances in the E$^{cm}_R=100-500$ keV range. Only two of these have been measured previously.
The strength of the E$^{cm}_R=482.7$ keV resonance was measured directly by \citet{Kuperus_1959}. To date, this is the lowest-lying, directly observed resonance in this reaction. For the E$^{cm}_R=439.1$ keV resonance, the proton partial width, $\Gamma_p$, was determined indirectly via the \hedreaction{30}{Si}{31}{P} reaction \citep{Vernotte_1990}. The other three resonances, E$^{cm}_R=144.3,\; 169.1,\; \text{and}\; 418.1$ keV, have not yet been measured. 
Their fractional contributions to the total rate are illustrated in the bottom panel of Figure~\ref{fig:38ar}. 
The E$^{cm}_R=418.1$ keV state is shown to have the largest contribution (50-100\%). A direct measurement of this resonance would therefore allow for a more precise determination of the reaction rate at these temperatures.
\section{Summary}
We reported here on a sensitivity study of the physical conditions that reproduce the abundance anomalies measured in NGC 2419. 
Using the methodology first presented in \citet{Iliadis_2016}, we obtained temperature and density combinations that simultaneously satisfy the observed abundances of Mg, Si, K, Ca, Sc, Ti, and V.
By varying each thermonuclear reaction rate within its probability distribution function, as listed in STARLIB, we found that the \pgreaction{37}{Ar}{38}{K}, \pgreaction{38}{Ar}{39}{K}, and \pgreaction{30}{Si}{31}{P} reactions are important in determining a more confined range of these conditions.  
\par
We then applied artificial systematic effects to the rates to explore the possibility that the acceptable temperature and density conditions were sensitive to unknown errors in the nuclear physics input.
We adjusted the rates from $1/10$ to $10$ times their recommended value and found this resulted in a substantive change in the acceptable solutions, with each rate changing the range of stellar temperatures required to produce the observed abundances.
In addition to the three reactions already identified, we also found that the \pgreaction{39}{K}{40}{Ca} reaction produced a marked change in these conditions.
To further test the influence that systematic effects could have, we then simulated two scenarios in which all four of these reactions rates were given artificial systematic effects.
In the first scenario, we adjusted each rate by a factor of either $5$ or $1/5$ to promote silicon synthesis and deter potassium production. In the second scenario, we did the opposite.
In comparing these two simulations we found that these minor rate adjustments changed the outcome of our simulations dramatically, leading either to significantly improved (first scenario) or worse (second scenario) constraints on the temperature and density conditions.
This suggests that new reaction rate measurements should lead to more robust predictions.
\par
Finally, we discussed the available nuclear structure data for the \pgreaction{30}{Si}{31}{P}, \pgreaction{37}{Ar}{38}{K}, \pgreaction{38}{Ar}{39}{K}, and \pgreaction{39}{K}{30}{Ca} reactions. Resonances important to the reaction rates at temperatures of $120-240$ MK have not been measured directly to date. Thus, we recommend their study as a means to refine the list of plausible polluter candidate sites in the cluster NGC 2419.

\acknowledgments
We would like to thank Sean Hunt and Lori Downen for their input and feedback.
This work was supported by the U.S. Department of Energy under Contract No. DE-FG02-97ER41041 and by NASA under the Astrophysics Theory Program Grant 14-ATP14-0007.

\newpage 

\bibliographystyle{aasjournal}
\bibliography{references}

\begin{thebibliography}{}
\expandafter\ifx\csname natexlab\endcsname\relax\def\natexlab#1{#1}\fi

\bibitem[{Angulo {et~al.}(1999)Angulo, Arnould, Rayet, Descouvemont, Baye,
  Leclercq-Willain, Coc, Barhoumi, Aguer, Rolfs, Kunz, Hammer, Mayer,
  Paradellis, Kossionides, Chronidou, Spyrou, Degl{\textquotesingle}Innocenti,
  Fiorentini, Ricci, Zavatarelli, Providencia, Wolters, Soares, Grama, Rahighi,
  Shotter, \& Rachti}]{Angulo_1999}
Angulo, C., Arnould, M., Rayet, M., {et~al.} 1999, Nuclear Physics A, 656, 3

\bibitem[{Cage {et~al.}(1971)Cage, Johnson, Kunz, \& Lind}]{Cage_1971}
Cage, M., Johnson, R., Kunz, P., \& Lind, D. 1971, Nuclear Physics A, 162, 657

\bibitem[{Cameron {et~al.}(2012)Cameron, Chen, Singh, \& Nica}]{Cameron_2012}
Cameron, J., Chen, J., Singh, B., \& Nica, N. 2012, Nuclear Data Sheets, 113,
  365

\bibitem[{Carretta {et~al.}(2010)Carretta, Bragaglia, Gratton, Recio-Blanco,
  Lucatello, D'Orazi, \& Cassisi}]{Carretta_2010}
Carretta, E., Bragaglia, A., Gratton, R.~G., {et~al.} 2010, Astronomy and
  Astrophysics, 516, A55

\bibitem[{Cohen {et~al.}(2011)Cohen, Huang, \& Kirby}]{Cohen_2011}
Cohen, J.~G., Huang, W., \& Kirby, E.~N. 2011, {ApJ}, 740, 60

\bibitem[{Cohen \& Kirby(2012)}]{Cohen_2012}
Cohen, J.~G., \& Kirby, E.~N. 2012, {ApJ}, 760, 86

\bibitem[{D'Ercole {et~al.}(2008)D'Ercole, Vesperini, D'Antona, McMillan, \&
  Recchi}]{D_Ercole_2008}
D'Ercole, A., Vesperini, E., D'Antona, F., McMillan, S. L.~W., \& Recchi, S.
  2008, Monthly Notices of the Royal Astronomical Society, 391, 825

\bibitem[{Doherty {et~al.}(2014)Doherty, Gil-Pons, Siess, Lattanzio, \&
  Lau}]{Doherty_2014}
Doherty, C.~L., Gil-Pons, P., Siess, L., Lattanzio, J.~C., \& Lau, H. H.~B.
  2014, Monthly Notices of the Royal Astronomical Society, 446, 2599

\bibitem[{Fuchs {et~al.}(1969)Fuchs, Grabisch, \& R{\"o}schert}]{Fuchs_1969}
Fuchs, H., Grabisch, K., \& R{\"o}schert, G. 1969, Nuclear Physics A, 129, 545

\bibitem[{Goriely {et~al.}(2008)Goriely, Hilaire, \& Koning}]{Goriely_2008}
Goriely, S., Hilaire, S., \& Koning, A.~J. 2008, Astronomy and Astrophysics,
  487, 767

\bibitem[{Goswami \& Prantzos(2000)}]{Goswami_2000}
Goswami, A., \& Prantzos, N. 2000, {A\&A}, 359, 191

\bibitem[{H{\"a}nninen(1984)}]{H_nninen_1984}
H{\"a}nninen, R. 1984, Nuclear Physics A, 420, 351

\bibitem[{Iliadis(2015)}]{Iliadis_Text}
Iliadis, C. 2015, Nuclear Physics of Stars, 2nd edn. (Weinheim: Wiley-VCH)

\bibitem[{Iliadis {et~al.}(2016)Iliadis, Karakas, Prantzos, Lattanzio, \&
  Doherty}]{Iliadis_2016}
Iliadis, C., Karakas, A.~I., Prantzos, N., Lattanzio, J.~C., \& Doherty, C.~L.
  2016, {ApJ}, 818, 98

\bibitem[{Iliadis {et~al.}(2015)Iliadis, Longland, Coc, Timmes, \&
  Champagne}]{Iliadis_2015}
Iliadis, C., Longland, R., Coc, A., Timmes, F.~X., \& Champagne, A.~E. 2015,
  Journal of Physics G: Nuclear and Particle Physics, 42, 034007

\bibitem[{Karakas(2010)}]{Karakas_2010}
Karakas, A.~I. 2010, Monthly Notices of the Royal Astronomical Society, 403,
  1413

\bibitem[{{Kikstra} {et~al.}(1990){Kikstra}, {Van Der Leun}, {Endt}, {Booten},
  {van Hees}, \& {Wolters}}]{Kikstra_1990}
{Kikstra}, S.~W., {Van Der Leun}, C., {Endt}, P.~M., {et~al.} 1990, Nuclear
  Physics A, 512, 425

\bibitem[{Kn{\"o}pfle {et~al.}(1974)Kn{\"o}pfle, Doll, Mairle, \&
  Wagner}]{Kn_pfle_1974}
Kn{\"o}pfle, K., Doll, P., Mairle, G., \& Wagner, G. 1974, Nuclear Physics A,
  233, 317

\bibitem[{Koning {et~al.}(2004)Koning, Hilaire, \& Duijvestijn}]{Konig_2004}
Koning, A.~J., Hilaire, S., \& Duijvestijn, M. 2004, TALYS: A Nuclear Reaction
  Program,  NRG-report 21297/04.62741/P

\bibitem[{Kuperus {et~al.}(1959)Kuperus, Smulders, \& Endt}]{Kuperus_1959}
Kuperus, J., Smulders, P., \& Endt, P. 1959, Physica, 25, 600

\bibitem[{Longland(2012)}]{Longland_2012}
Longland, R. 2012, Astronomy {\&} Astrophysics, 548, A30

\bibitem[{{Longland}(2017)}]{Longland_2016}
{Longland}, R. 2017, {in preparation}

\bibitem[{Longland {et~al.}(2010)Longland, Iliadis, Champagne, Newton, Ugalde,
  Coc, \& Fitzgerald}]{Longland_2010}
Longland, R., Iliadis, C., Champagne, A., {et~al.} 2010, Nuclear Physics A,
  841, 1

\bibitem[{Moreh {et~al.}(1988)Moreh, Sandefur, Sellyey, Sutton, \&
  Wildenthal}]{Moreh_1988}
Moreh, R., Sandefur, W.~M., Sellyey, W.~C., Sutton, D.~C., \& Wildenthal, B.~H.
  1988, Phys. Rev. C, 37, 2428

\bibitem[{Mucciarelli {et~al.}(2012)Mucciarelli, Bellazzini, Ibata, Merle,
  Chapman, Dalessandro, \& Sollima}]{Mucciarelli_2012}
Mucciarelli, A., Bellazzini, M., Ibata, R., {et~al.} 2012, Monthly Notices of
  the Royal Astronomical Society, 426, 2889

\bibitem[{Nolan {et~al.}(1981)Nolan, Al-Naser, Behbehani, Butler, Green, James,
  Lister, Rammo, Sharpey-Schafer, \& Sheppard}]{Nolan_1981}
Nolan, P.~J., Al-Naser, A.~M., Behbehani, A.~H., {et~al.} 1981, J. Phys. G:
  Nucl. Phys., 7, 189

\bibitem[{Piotto {et~al.}(2007)Piotto, Bedin, Anderson, King, Cassisi, Milone,
  Villanova, Pietrinferni, \& Renzini}]{Piotto_2007}
Piotto, G., Bedin, L.~R., Anderson, J., {et~al.} 2007, {ApJ}, 661, L53

\bibitem[{Rolfs \& Rodney(1988)}]{Rolfs_1988}
Rolfs, C., \& Rodney, W. 1988, Cauldrons in the Cosmos: Nuclear Astrophysics
  (University of Chicago Press)

\bibitem[{Sallaska {et~al.}(2013)Sallaska, Iliadis, Champange, Goriely,
  Starrfield, \& Timmes}]{Sallaska_2013}
Sallaska, A.~L., Iliadis, C., Champange, A.~E., {et~al.} 2013, ApJS, 207, 18

\bibitem[{Seth {et~al.}(1967)Seth, Biggerstaff, Miller, \&
  Satchler}]{Seth_1967}
Seth, K.~K., Biggerstaff, J.~A., Miller, P.~D., \& Satchler, G.~R. 1967,
  Physical Review, 164, 1450

\bibitem[{Ventura {et~al.}(2012)Ventura, D'Antona, Criscienzo, Carini,
  D'Ercole, \& vesperini}]{Ventura_2012}
Ventura, P., D'Antona, F., Criscienzo, M.~D., {et~al.} 2012, {ApJ}, 761, L30

\bibitem[{Vernotte {et~al.}(1990)Vernotte, Khendriche, Berrier-Ronsin,
  Grafeuille, Kalifa, Rotbard, Tamisier, \& Wildenthal}]{Vernotte_1990}
Vernotte, J., Khendriche, A., Berrier-Ronsin, G., {et~al.} 1990, Phys. Rev. C,
  41, 1956

\bibitem[{Villanova {et~al.}(2007)Villanova, Piotto, King, Anderson, Bedin,
  Gratton, Cassisi, Momany, Bellini, Cool, Recio-Blanco, \&
  Renzini}]{Villanova_2007}
Villanova, S., Piotto, G., King, I.~R., {et~al.} 2007, {ApJ}, 663, 296

\end{thebibliography}

\end{document}